\documentclass[11pt]{article}
\usepackage{geometry}                
\usepackage{graphicx}
\usepackage{amssymb}
\usepackage{amsmath}
\usepackage{amsthm}
\usepackage{epstopdf}
\usepackage{subfigure}

\usepackage{gensymb}

\newtheorem{question}{Question}
\newtheorem{answer}{Answer}

\title{On exploration of geometrically constrained space by medicinal leeches \emph{Hirudo verbana}}
\author{Andrew Adamatzky \\
Unconventional Computing Centre and Bristol Robotics Lab,\\
University of the West of England, UK}

\begin{document}
\maketitle

\begin{abstract}
\noindent
Leeches are fascinating creatures:  they have simple  modular nervous circuitry m yet exhibit a rich spectrum of behavioural modes. Leeches could be ideal blue-prints for designing flexible soft robots which are modular, multi-functional, fault-tolerant, easy to control, capable for navigating using optical, mechanical and chemical sensorial inputs, 
have autonomous inter-segmental coordination and adaptive decision-making. With future designs of leech-robots in mind we study how leeches behave in geometrically constrained spaces. Core results of the paper deal with leeches exploring a row of rooms arranged along a narrow corridor. In laboratory experiments we find that rooms closer to 
ends of the corridor are explored by leeches more often than rooms in the middle of the corridor. Also, in series of scoping experiments, we evaluate leeches capabilities to navigating in mazes towards sources of vibration and chemo-attraction. We believe our results lay foundation for future developments of robots mimicking behaviour of leeches. 

\vspace{0.5cm}

\noindent
\emph{Keywords:}  medicinal leech, space exploration, maze navigation, bio-robotics
\end{abstract}

\section{Introduction}

Creatures explore space around them in a search for nutrients, mating partners, or more favourable environmental conditions: exact search strategies vary amongst species yet  common features include regular scanning with sensors and following gradients towards attractants or away from repellents~\cite{birke1983some, atema1996eddy, murphy1978practical, ellen1982spatial, wood1983exploration, wood1990significance, krohn1994behaviour, wood1993inquisitive, armstrong1997spatial, mettke2002significance, hutt1966exploration, macri2002risk,  qin2007maze}. 
Principles, mechanisms and strategies of the explorative and foraging behaviour were extensively used and adopted 
in bio-inspired algorithms of computational optimisation~\cite{dorigo1996ant, dorigo2006ant, dorigo2010ant, yang2005engineering, yang2010nature, yang2009firefly, yang2010new, adamatzky2010physarum, jones2014computation, zhu2013amoeba, tsompanas2013evolving, tsompanas2014physarum}, software governing navigation of mobile bio-inspired robots~\cite{yamauchi1998frontier, grabowski2000heterogeneous, castillo2012comparative, sheynikhovich2005spatial, pfeifer2012challenges, milford2007spatial, cuperlier2007neurobiologically, beni1993swarm}
and embedded unconventional robotic controllers~\cite{melhuish2001biologically, adamatzky2003experimental, adamatzky2004experimental, adamatzky2009artificial}.

Designing robots with worm-like bodies became now one of the most hot topics in bio-inspired robotics. This is because worm and snake like robots have advantages of high flexibility, modularity, multi-functionality,  distributed adaptability, relative ease of controllability, allow for multiple  engineering implementations, and can be manufactured from a wide range of materials.  A key feature of worm-robots is that they are capable for accessing places other, wheeled or walking, robots can not access. The feature becomes of  a particular importance when the main task of the robots is to implement search and rescue operations~\cite{gorges12014cascaded}. Research in worm-robots produced a great variety of high-impact results. Representative examples include climbing worm robots controlled by oscillatory networks~\cite{wang2008cpg}, worm robots propagating in flexible  environments~\cite{zarrouk2012conditions}, neumatic flexible robot prototype for pipes inspection~\cite{bertetto2001pipe}, earthworm inspired robot~\cite{zarrouk2010analysis}, robots capable for drilling and excavation applications~\cite{kubota2007earth}, robots with magnetizable elastic 
body~\cite{zimmermann2007deformable}, ring-worm like robot~\cite{arena1999reaction}, robots propelled 
by peristaltic motion with artificial muscles~\cite{boxerbaum2010new, arena2002electro}, Oligochaetes-inspired robots~\cite{seok2013meshworm}, and elastic robots for endoscopy \cite{manwell2014elastic}.

Despite such immense progress in the design of worm-inspired robots the issues of space exploration by robots, especially in the situations, when there are no attracting or repelling stimuli, remain largely not addressed. What bio-inspired robots do when they have not a clue of what they are searching for. This is why we decided to conduct a series of experiments in uncovering principal patterns of geometrically constrained space exploration by leeches. Why leeches? Why geometrically constrained space? The answers are below.  

The leech \emph{Hirudo medicinal} and its South European analog \emph{Hirudo verbana} are amongst most common living creatures explored in laboratory conditions. The leeches has relatively simple nervous system yet they exhibit a wide spectrum of complex behaviour, sometimes even sophisticated traits of parental care~\cite{kutschera1986leech}, mapped to identified neural circuits.  A leech is amongst most popular living substrates for modelling nervous system and locomotion control~\cite{KristanJr1977191, lockery1993computational, campos2007temporal, gaudry2010feeding, crisp2012mechanisms, kristan2005neuronal, friesen1993mechanisms, brodfuehrer1993effect, lockery1993lower}, modulating behaviour of neuro-mediators~\cite{zaccardi2004sensitization, alkatout2007serotonin, gerry2012serotonin},  developmental processes in complex neuronal circuits~\cite{kristan2000development},  and mathematical and computers models of circuits responsible for regular pattern generations~\cite{taylor2000model, zheng2007systems, pearce1988model, buono2004mathematical}. The leeches' neural networks are simple yet efficient, they are equivalent in their computational power to basic perceptrons~\cite{lockery1993computational}.

Leeches are ideal inspirations for amphibious soft robots, capable for reaching spaces not accessible by other devices.\footnote{While writing about `devices' we thought it is interesting to note, that in the UK medicinal leeches have been classified as a medicinal product but in the US they are classified as a medical device ~\cite{taneja2011national}.} 
The reasons are following. A leech has a modular structure, its body segmentation along the anterior-posterior axis is very convenient for robotic implementations. A single segment of the leech's body contains isolated ganglion. It is capable for exhibiting swimming activity even when the segment has been neurally isolated from the rest of the leech's body for one-two weeks~\cite{KristanJr1977191}. A spectrum of leeches' behaviour traits is well classified, and therefore can be adopted with minimal efforts in amphibious~\cite{crespi2005swimming, crespi2004amphibious, yang2007preliminary, crespi2005amphibot, yang2008body, yu2009amphibious} robotic devices:  a leech positions itself at the water surface in resting state; the leech swims towards the source of a mechanical or optical stimulation; 
the leech stops swimming when comes into contact with any geometrical surface; then, the leech explores the 
surface by crawling; when a leech finds a warm (37-40$^o$C) region the leech bites~\cite{dickinson1984feeding}. 
Moreover, a behaviour of a leech is context based: the leech can respond to constant sensorial inputs with variable motor outputs~\cite{brodfuehrer2001identified}.

We focus on geometrically constrained spaces  because behaviour of leeches in uniform spaces is analysed in full details:  in empty space leeches wander around uniformly with no preferential direction or location~\cite{garcia2005statistics}. Little known about leeches behavioural patterns in complex geometries (apart of the fact that leeches show positive stigmotaxis which lead them to crawl into body cavities for feeding and under logs and stones when fed), and, to our knowledge, no results are published on a role of geometries of environmental shapes in patterning the leeches's behaviour. 

The paper is structured as follows. Section~\ref{methods} introduces the problem and experimental design used to study behaviour of leeches in a corridor with a raw of rooms. Statistical analysis of the behaviour and finite state machine model are presented in Sect.~\ref{results}. Scoping experiments on navigation of leeches in complex geometrically spaces --- mazes, and ways of potential further developments are outlined in Sect.~\ref{furtherdevelopments}. Final touching remarks are done in Sect.~\ref{discussion}.

\section{Methods}
\label{methods}

\begin{question}
Let you be in a hotel, standing in a long corridor with a row of rooms on one side. 
The hotel soon to be invaded by giant leeches. 
What room  you must hide in to decrease your chances of being bitten by the leeches?
\end{question}

\begin{figure}[!tbp]
\centering
\subfigure[]{\includegraphics[width=0.49\textwidth]{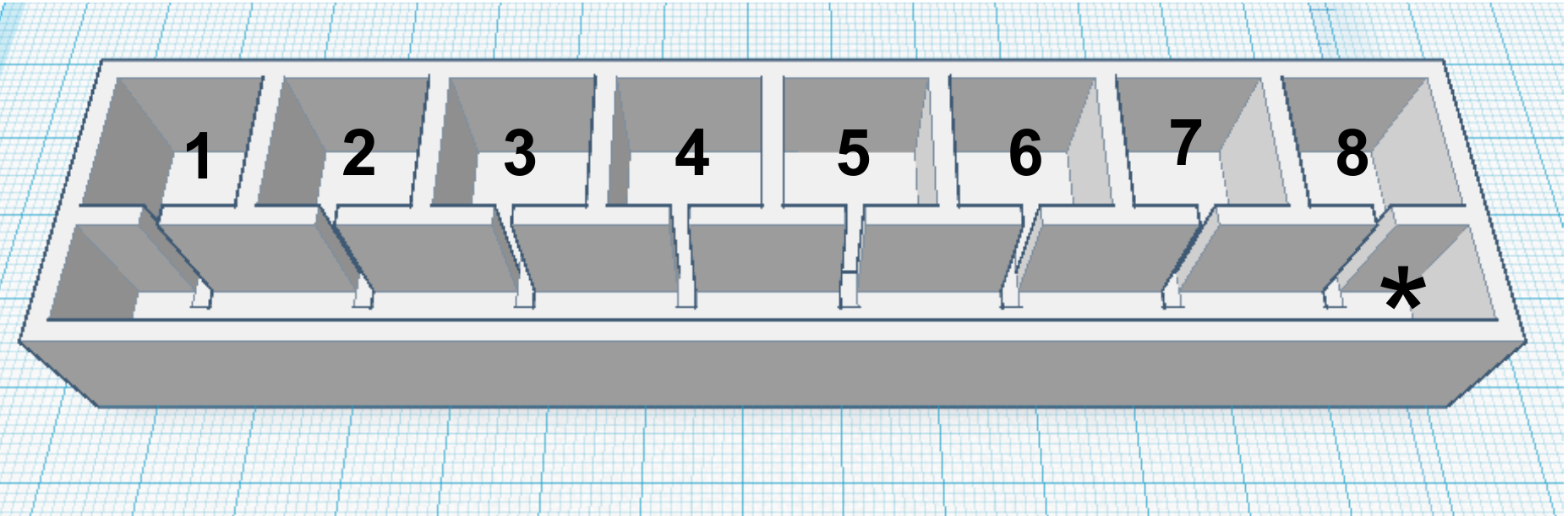}}
\subfigure[43~s]{\includegraphics[width=0.49\textwidth]{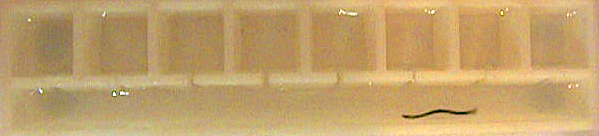}}
\subfigure[70~s]{\includegraphics[width=0.49\textwidth]{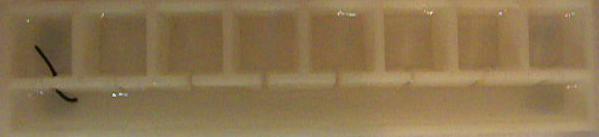}}
\subfigure[110~s]{\includegraphics[width=0.49\textwidth]{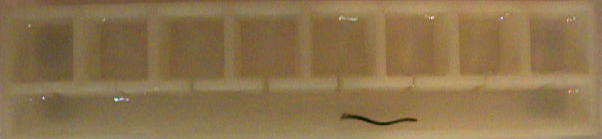}}
\subfigure[159~s]{\includegraphics[width=0.49\textwidth]{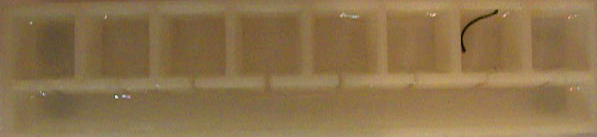}}
\subfigure[178~s]{\includegraphics[width=0.49\textwidth]{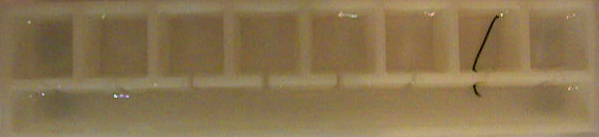}}
\subfigure[208~s]{\includegraphics[width=0.49\textwidth]{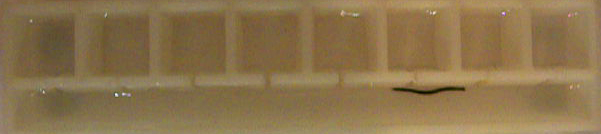}}
\subfigure[250~s]{\includegraphics[width=0.49\textwidth]{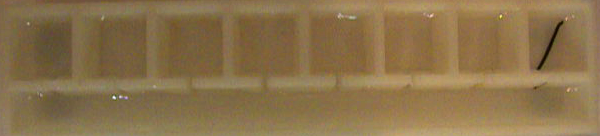}}
\subfigure[279~s]{\includegraphics[width=0.49\textwidth]{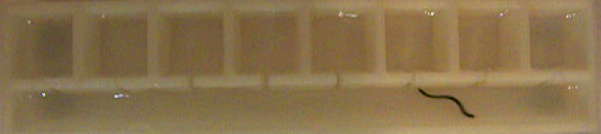}}
\subfigure[378~s]{\includegraphics[width=0.49\textwidth]{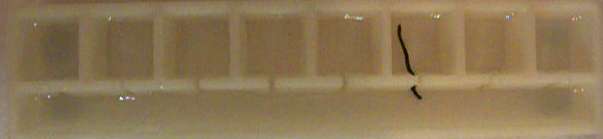}}
\caption{Experimental setup. (a)~Scheme of the template used: initial position of a leech is shown by star, rooms are numbered as in the picture. (b)~Photographs of a sample experiment; time shown is the time elapsed form the moment the leech was placed in the template.}
\label{setup}
\end{figure}

We used three weeks old leeches  \emph{Hirudo verbana} obtained from Biopharm Leeches (Hendy, Carmarthenshire SA4 0X, UK). Leeches varied in size from 15 to 20~cm length in elongated state and 1-2~mm width. Leeches awaiting experiments were kept in securely covered, yet with air access, glass containers in a dechlorinated water away from direct sunlight. As per recommendation~\cite{taneja2011national} leeches were kept in a cool, c. 15$\degree$C, environment to lessen their needs for feeding and to enhance their performance in exploration of experimental templates. The water was refreshed every other day. When moving leeches between storage containers and experimental templates we used non-serrated forceps. The template was printed from polylactic acid thermoelastic polyester (Fig.~\ref{setup}a) with the following dimensions: length 138~mm, width 31~mm, height 17~mm, wall where 2~mm in width everywhere. The internal space was subdivided by eight rooms and a corridor. The corridor width was 10~mm.  Each room had dimensions 15$\times$15~mm with 2~mm opening into the corridor. Experimental template was  cleaned to remove any substances with odour or state that might affect behaviour of leeches in the templates. The template was filled with dechlorinated water, depth 10-15~mm; the leeches  were able to swim if they wanted to. The dechlorinated water in the experimental templates was changed after each experiment to prevent metabolites and ions released by leeches to affect behaviour of their successors. The template was illuminated by LED lamp, illumination level at the bottom of the template
was 37 LUX. No sharp gradients of optical, chemical or electoral stimuli were allowed; the only stimulation occurred was mechanical ones when leeches come into contact with walls of the templates. Experiments were conducted in a room temperature of 20$\degree$C. 

We conducted 40 experiments using 20 leeches; each leech was used twice with at least 24 h interval between experiments.  In each trial a leech was placed at the right end of the corridor as marked by star in Fig.~\ref{setup}a. 

The experiments were recorded on Coolpix P90 digital camera,  640 $\times$ 480 pixels frame size and 25 frames per second speed. Each video was recorded for c. 25--30~min. The videos were analysed by in-house software written in Processing, as follows. For every second of video we extracted coordinates of pixels with colour values less than 30-50 in RGB mode (exact threshold was quickly adjusted for each video). Such pixel represented body of the leach. Their coordinates with time tags were stored for further analyses. Configurations of leeches exploring the template were converted to overlay images with colours (Fig.~\ref{examplestimecoloured}) as follows. For any trial/video a duration of recording is normalised to the interval [0,1] and then mapped to a colour scale 
$(00B)$ $\rightarrow$  $(0BG)$ $\rightarrow$ $(RG0)$ $\rightarrow$ $(R00)$, i.e. the blue pixel represent leech at the beginning of experiment and red pixel at the end of experiment.

\section{Results}
\label{results}

Exemplar snapshots of an experiment are show in Fig.~\ref{setup}b--j. In this particular experiment a leech released in the right end of the corridor crawls along the corridor towards its left end (Fig.~\ref{setup}b). On reaching the end of the corridor the leech starts to explore room 1 (Fig.~\ref{setup}c): this is a positive stigmotaxis, coming into contact with walls, especially in corners, evokes the explorative behaviour. On leaving room 1 the leech crawls along the corridor  
(Fig.~\ref{setup}d) and  then enters room 7 where she remains (exploring the corners of the room) for half a minute
(Fig.~\ref{setup}ef). The leech's further activities involved swimming along corridor (Fig.~\ref{setup}g) and exploration 
of rooms 8   (Fig.~\ref{setup}h) and 6 (Fig.~\ref{setup}ij).

\begin{figure}[!tbp]
\centering
\subfigure[]{\includegraphics[width=0.49\textwidth]{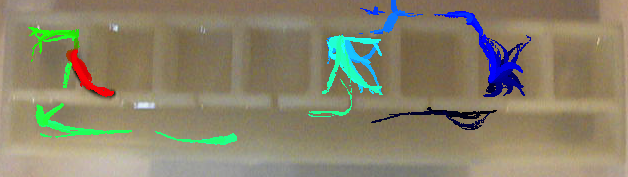}}
\subfigure[]{\includegraphics[width=0.49\textwidth]{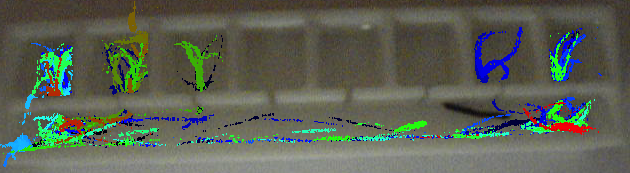}}
\subfigure[]{\includegraphics[width=0.49\textwidth]{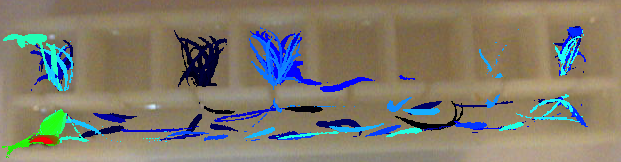}}
\subfigure[]{\includegraphics[width=0.49\textwidth]{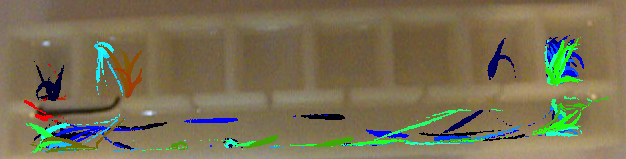}}
\subfigure[]{\includegraphics[width=0.49\textwidth]{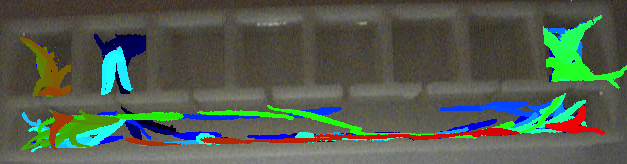}}
\subfigure[]{\includegraphics[width=0.49\textwidth]{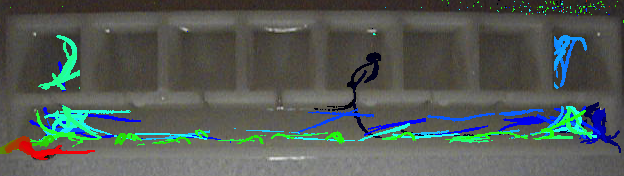}}
\subfigure[]{\includegraphics[width=0.49\textwidth]{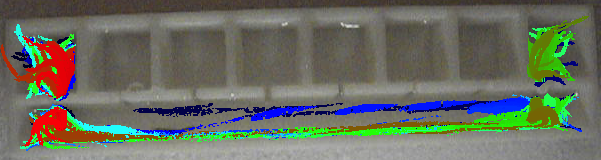}}
\subfigure[]{\includegraphics[width=0.49\textwidth]{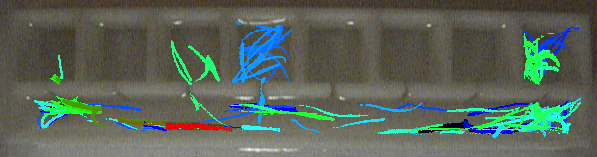}}
\caption{Overlay of patterns of the leech exploring the template for eight exemplary, 
and most typical, trials.}
\label{examplestimecoloured}
\end{figure}

Representatives examples of leeches exploratory trajectories are shown in Fig.~\ref{examplestimecoloured}.  In experiment Fig.~\ref{examplestimecoloured}a a leech crawls into room 7, then gets out of the room and moves into room 5. Further the leech crawls towards left end of the corridor and explores room 1. While exploring the room 1 the leech raises above water surface and leans to the room 2.  During experiment shown in 
 Fig.~\ref{examplestimecoloured}b a leech visits room 7 then moves along the corridor back and forth and visits 
 rooms 8 and 3. The leech visits room 7 once, rooms 8 and 3 twice, room 2 twice, and room 1 at least four times. 
 Experiments illustrated in  Fig.~\ref{examplestimecoloured}c-h demonstrate that in majority of trials leach definitely 
 visits rooms 1 and 8, and with lesser chances rooms 2, 3 and 6, 7 and even more rarely rooms 4 and 5. 

\begin{figure}[!tbp]
\centering
\subfigure[]{\includegraphics[width=0.9\textwidth]{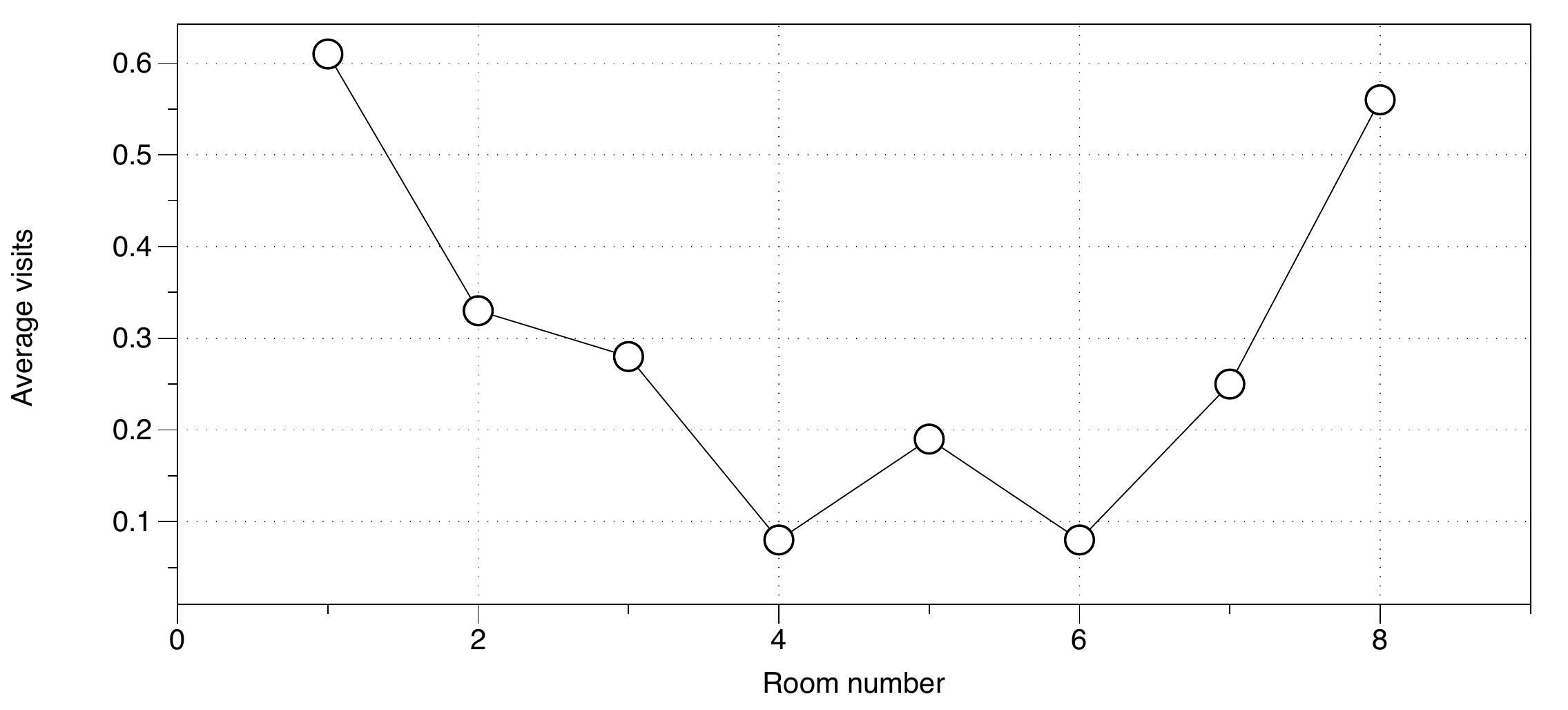}}
\subfigure[]{\includegraphics[width=0.9\textwidth]{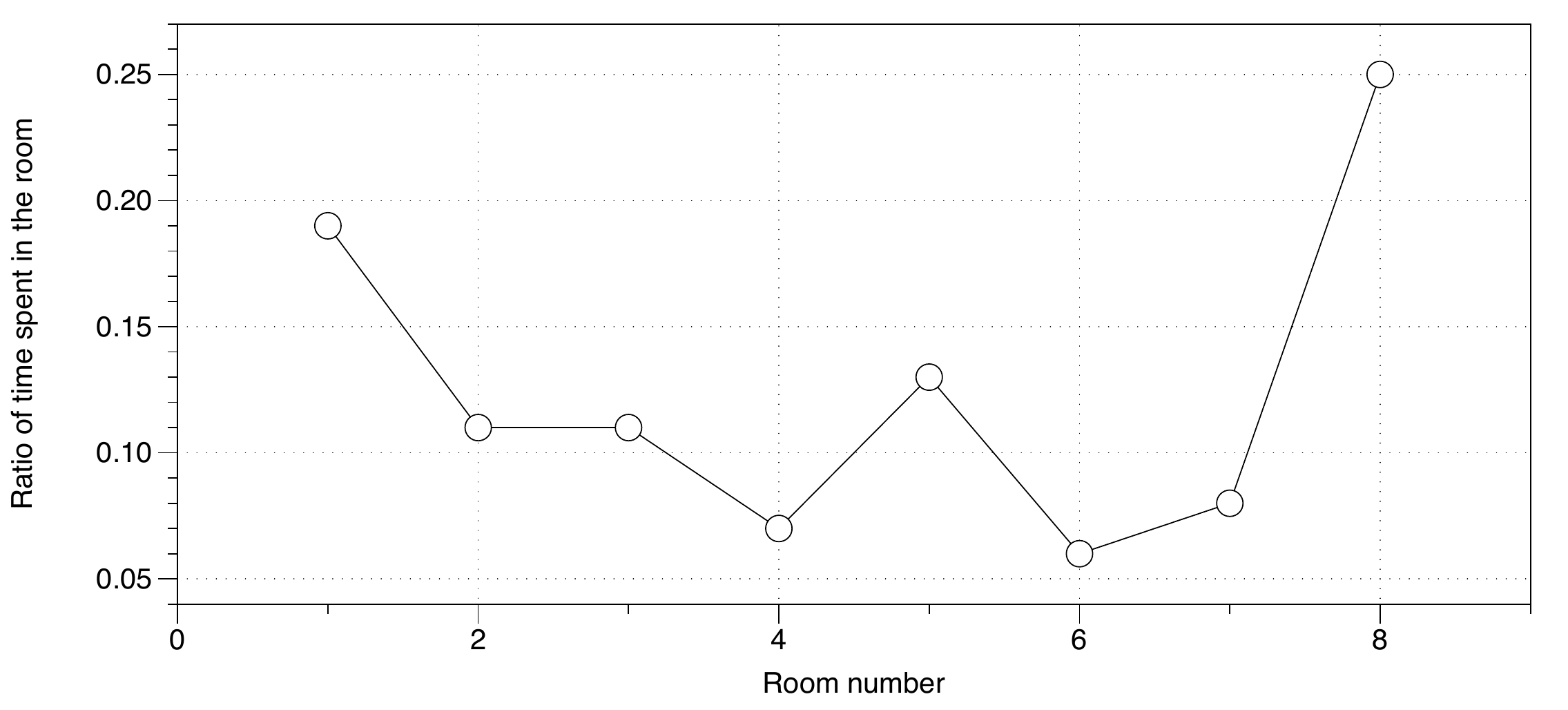}}
\caption{Statistics of the template exploration by leeches. (a)~Frequency of room visits. 
(b)~Ratio of time spent in each particular room. }
\label{statistics}
\end{figure}

Statistical evaluation of room exploration is shown in Fig.~\ref{statistics}. In majority of trials leeches explore rooms at the ends of the corridor, room 1 and 8 (Fig.~\ref{statistics}a). Then chances of a room being visited by a leeche are inversely proportionally  to a distance from closest end of the corridor and the room. Amount of the time a leech spends in a room is also inversely proportional to a distance between the room and the closest end of the corridor 
(Fig.~\ref{statistics}b). For example, a leech typically spends five times more time in the room 8 than in the room 6, or three times more in the room 1 than in the room 4. Slight apparent increase of time spent by leeches in room 5 might be due to unknown reasons or noise in experimental evaluations. 
 
 \begin{answer}
 Being in a corridor with a row of rooms to be invaded by leeches the best way to escape the leeches is to hide in 
 a room in the middle of the row. 
 \end{answer}
 
All four forms of leech behaviour,  as classified in~\cite{garcia2005statistics}, were observed: stationary/still (interior and posterior sucker attached to the bottom of the template and leech does not make any motion), swimming (undulatory movement of the entire body),   crawling (alternating steps of elongation and contraction with posterior and anterior suckers attached in turns),  exploratory (irregular oscillations of the head and anterior part of the body with the posterior sucker attached).\footnote{Definitions of the types are taken almost verbatim from~\cite{garcia2005statistics}}. Indeed feeding is a dominant pattern of leech behaviour~\cite{misell1998behavioral} however leeches did not have a chance to feed in our experiments, therefore we did not include feeding in out analysis. Swimming form of behaviour was quite rare: only eight of 40 experiments leeches were observed swimming along the corridor. Crawling, exploratory and stationary/still modes were most common. 
 
 \begin{table}[!tbp]
 \caption{Transitions probabilities of leech automaton ${\cal L}$.}
 \begin{tabular}{l||l|lll}
 	&   		&       		&    $s^{t+1}$ &  \\ \hline \hline
 	&    		& Still 		& Crawl 			& Explore \\ \hline
$s^t$ &  Still     	&  $1-p_1$   &  $\frac{1}{2}p_1$       	&  $\frac{1}{2}p_1$    \\
 	&  Crawl  	&  $p_2$     	&  $ (1-m^t) (1-p_2)+(1 - p_3)$       & $ m^t (1-p_2)+p_3$    \\
 	&  Explore  	&  $p_2$    	&  $ (1-m^t) (1-p_2)$       &   $m^t (1-p_2) $   \\
 \end{tabular}
 \label{transitions}
 \end{table}

Leeches make their decisions sequentially~\cite{esch2002evidence}, thus we can propose a leech automaton, a finite state machine equipped with mechanoreceptor and a timer:
${\cal L}=\langle  {\mathbf Q},  {\mathbf M},  {\mathbf  T} \rangle$, where  
${\mathbf Q}$=$\{$ Still, Crawl, Explore $\}$ is a set of leech's behavioural mods;  
 ${\mathbf M}$= $\{ 0, 1 \}$ is a set of of states of mechanoreceptors, where `1' means the mechanoreceptor detects
 obstacles and `0' no obstacle is detected; ${\mathbf  T}$ is set of timer states.  
 
 At each iteration  the  automaton ${\cal L}$ is represented by its internal state $s^t \in {\mathbf Q}$, 
 state of its mechanoreceptors $m^t \in {\mathbf M}$ and its timer $t \in {\mathbf  T}$.  
The timer state $t$ is reset, $t \leftarrow 0$, when automaton ${\cal L}$ undergoes transitions: 
$\{$Crawl, Explore$\}$$\rightarrow$Still or Still$\rightarrow$$\{$Crawl, Explore$\}$, otherwise 
timer state increments $t \leftarrow t+1$. The timers imitates spontaneous transitions between inactive and active states: ``hungry leeches often initiate swimming bouts which are interposed between longer periods of quiescent resting at the water surface''~\cite{dickinson1984feeding}.

 Conditions of the transitions between the internal states ${\mathbf Q}$, expressed in terms of probabilities are  shown in Tab.~\ref{transitions}. Probabilities of transitions are as follows 
 $p_1 = (\tau_s - t +1)^{-1}$,  
 $p_2 = (\tau_a - t +1)^{-1}$, 
 where  $\tau_s$ and $\tau_a$ are time intervals a leech spends in still or active (crawling or exploring) modes, respectively. 
 Value of $\tau_a$ observed in our experiments varies from 10~min to 20~min, value of $\tau_s$ will be estimated in further studies.  The probability $p_3$ of switching from crawling mode to explorative mode is derived from our experimental data  (Fig.~\ref{statistics}): if $x$ is a distance from a room to the closest end of the corridor, than a probability that a leech visits the room at least once is $p_3=0.35 \cdot x^{-0.82}$.  The structure of the leech automaton ${\cal L}$ proposed conforms to results  by Garcia-Perez et al.~\cite{garcia2005statistics} that the leech's behavioural changes are well described by Markov processes, with transition rates controlled by the firings of command-like neurons.

\section{Discussion: Further developments and scoping experiments}
\label{furtherdevelopments}

\begin{figure}[!tbp]
\centering
\subfigure[]{\includegraphics[width=0.49\textwidth]{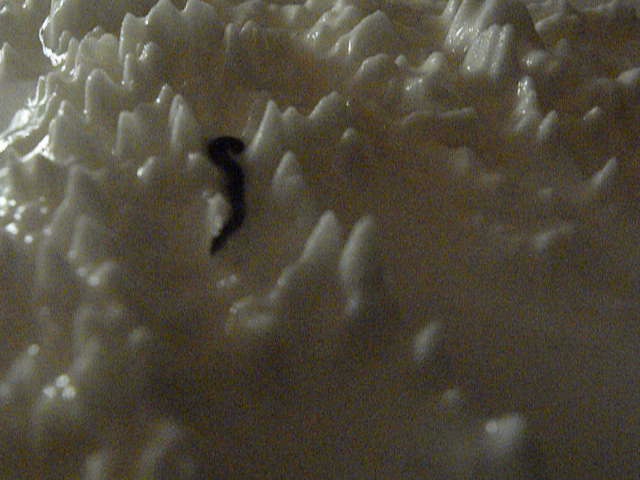}}
\subfigure[]{\includegraphics[width=0.49\textwidth]{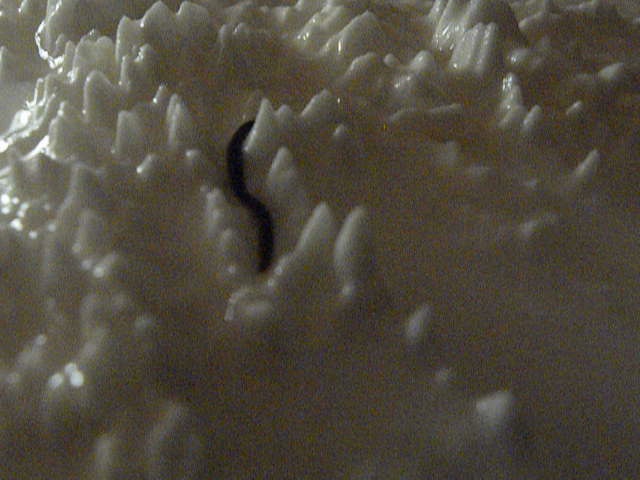}}
\subfigure[]{\includegraphics[width=0.49\textwidth]{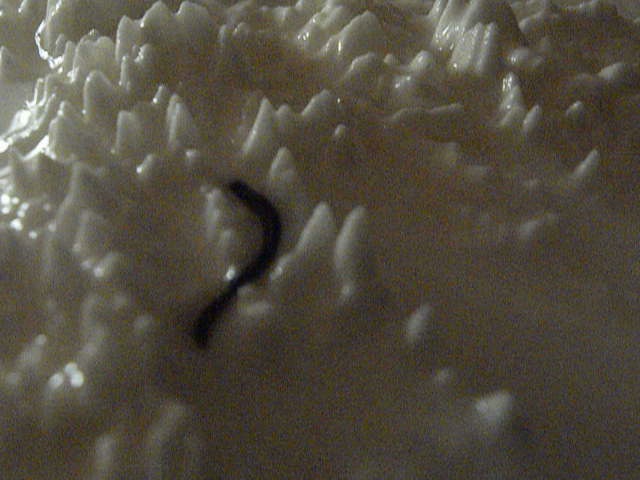}}
\subfigure[]{\includegraphics[width=0.49\textwidth]{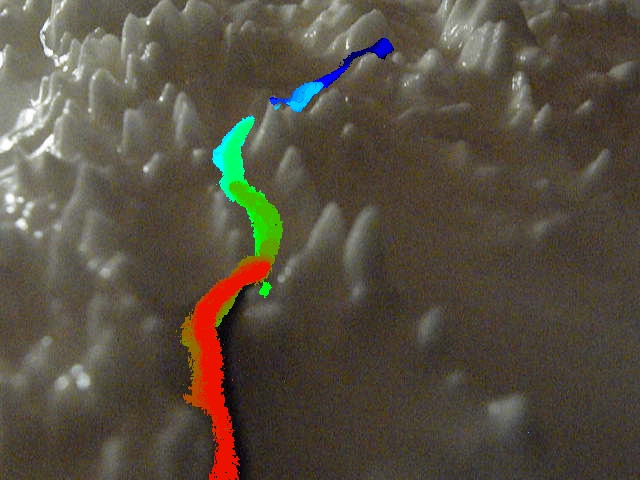}}
\caption{Photographs (abc) and an exemplar time-snapshots (d) of leeches navigating around three-dimensional terrain of Balkans. Time is encoded from blue to green to red.}
\label{balkans1}
\end{figure}

\begin{figure}[!tbp]
\centering
\subfigure[]{\includegraphics[width=0.49\textwidth]{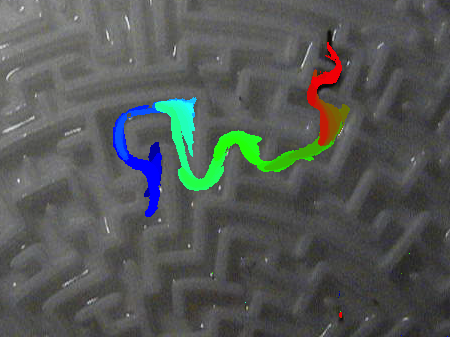}}
\subfigure[]{\includegraphics[width=0.49\textwidth]{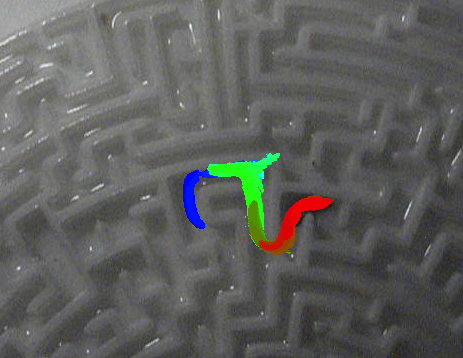}}
\caption{Time-snapshots of leeches propagating in labyrinth. Time is encoded from blue to green to red. }
\label{maze}
\end{figure}

To evaluate how well leeches propagate in geometrically constrained space, which are a bit more complex than
the row of rooms, we used two templates of mazes (15~cm diameter round maze printed in nylon and 
7$\times$7~cm square acrylic maze) and a three-dimensional template of Balkans (printed from nylon, 20$\times$14$\times$7~cm). 

Results were promising. In three-dimensional template of Balkans, just slightly (c. 3-5~mm) covered by dechlorinated water leeches navigated around elevations, see examples in Fig.~\ref{balkans1}. In a maze the leeches moved along the channels (Fig.~\ref{maze}). As shown in Fig.~\ref{maze}, in many case a leech explores corners of the maze channels or even checks the space above the wall in the corner before making a turn. We concluded that long term experimenting with leeches in the maze would be unreasonable because the leeches frequently attempt, and often succeed, in  making short-cuts by crawling over walls between the maze channels. Thus below we discuss a bit of our findings in potential ways of guiding leeches towards targets in complex geometries.  

In experiments described in Sect.~\ref{results} there were no attractants or repellents because we aimed to study how only geometrical constrains affect leeches behaviour. Leeches, such \emph{Hirudo verbana} determine prey location by sensing water disturbances using both visual and tactile sensors~\cite{young1981responses, dickinson1984feeding,harley2013discontinuous}.  A typical scenario would be as follows. A leech is resting near water surface or on the bottom of a pond. Animals enter the pond for drinking, their moving tongues generate waves which are projected as intermittent stripes of shadows and light on the bottom of the pond. Leeches swim towards the surface. Then they start utilising feedback from their mechanical sensors and propagate towards the source of waves. Exact place of the leeches attachment to their prey is determined by temperature and, possibly, chemical gradients. 
Leeches respond to mechanical stimulation from surface generated waves but orienting towards the source of the waves using their mechanical receptors systems; visual stimuli evoke less accurate responses~\cite{carlton1993comparison}.  Experiments presented in paper \cite{harley2013discontinuous} demonstrate that leeches do not move towards source of tactile stimulation ballistically but under a guidance of continuous sensory input.

To check ability of leeches to navigate in complex geometries guided by attractants we conducted a series of scoping experiments, as follows.

\begin{figure}[!tbp]
\centering
\subfigure[]{\includegraphics[width=0.75\textwidth]{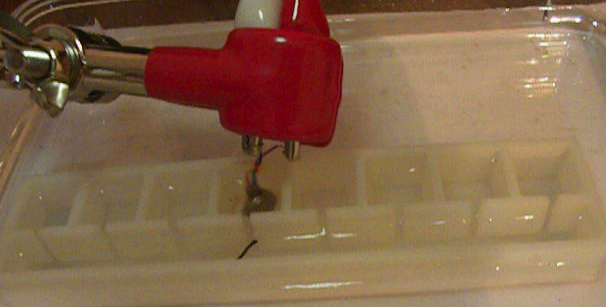}}
\subfigure[]{\includegraphics[width=0.75\textwidth]{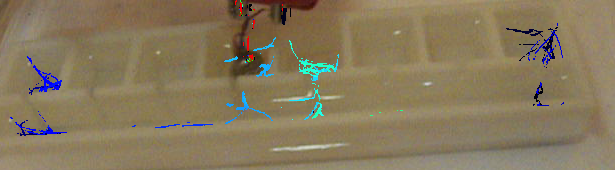}}
\subfigure[]{\includegraphics[width=0.8\textwidth]{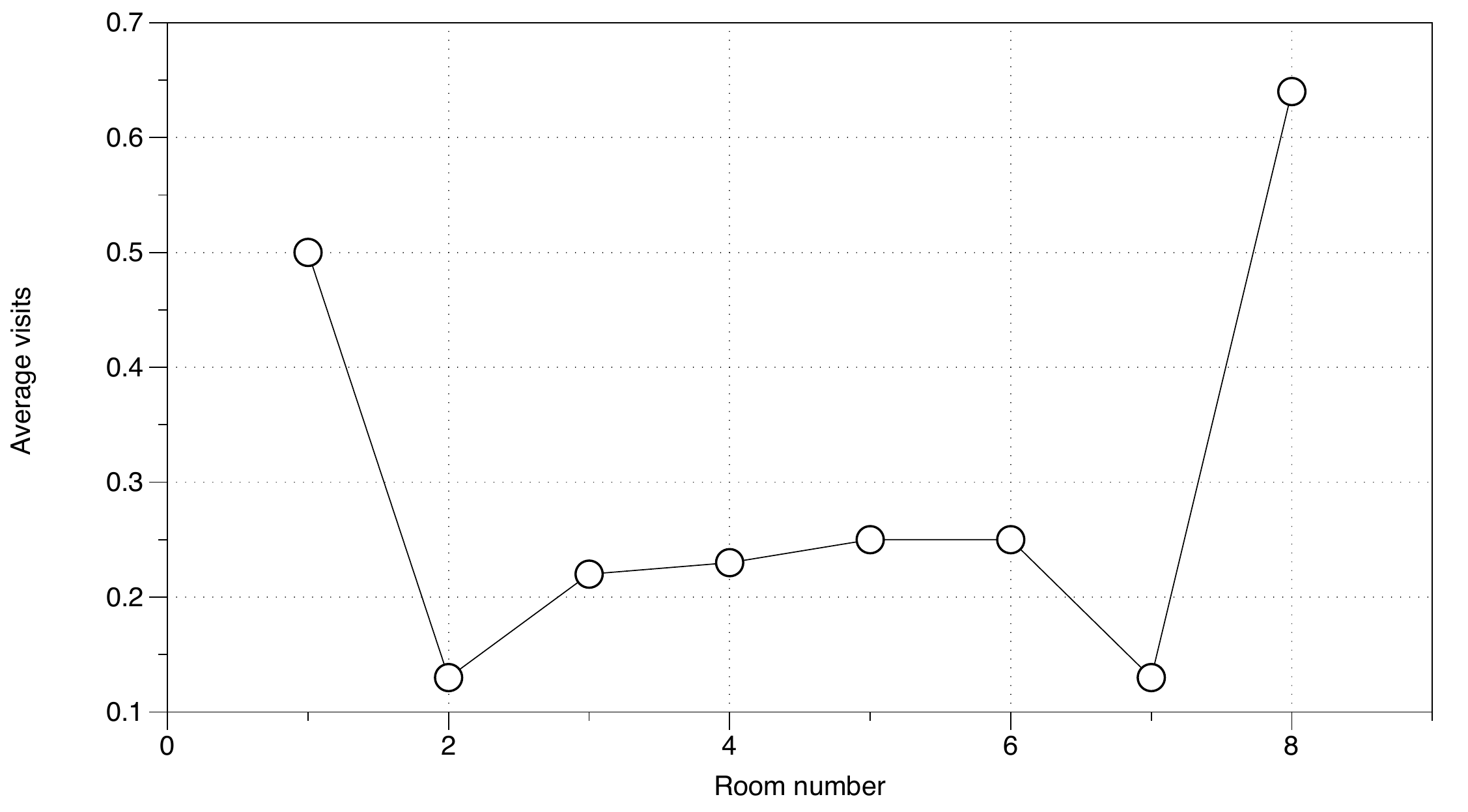}}
\caption{Guiding a leech by waves. (a)~Photo of experimental setup: the vibrating motor is suspend above room 4 and slightly immersed in the water, leech is entering the room 4. (b)~Time-snapshots of the leech exploratory activities.   (c )~Statistics of rooms exploration collected in eight experiments. }
\label{vibrationcorridor}
\end{figure}

\begin{figure}[!tbp]
\centering
\subfigure[]{\includegraphics[width=0.49\textwidth]{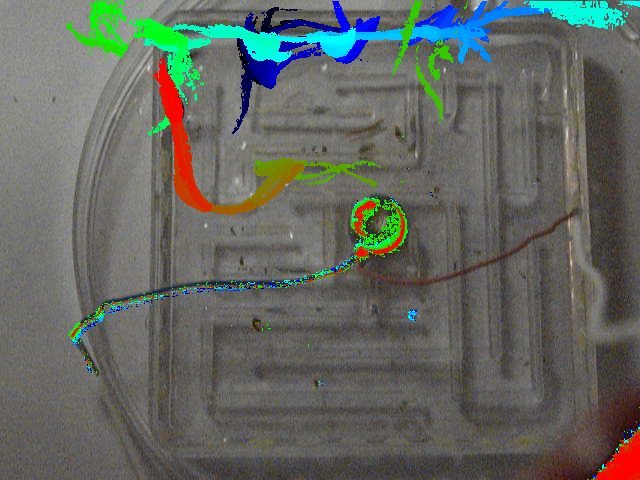}}
\subfigure[]{\includegraphics[width=0.49\textwidth]{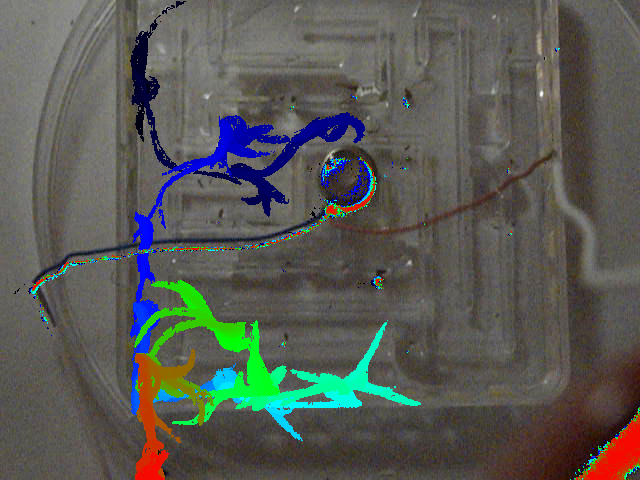}}
\caption{Two exemplary time-snapshots of a leech exploring small maze where a vibrating motor 
was placed in a central chamber.}
\label{mazevibration}
\end{figure}

To induce surface waves we used Pico Vibe\texttrademark{ }  Vibration Motor  (Precision Microdrives, London, UK). The motors has  8~mm body diameter, 2.15~mm body length, 3~V rated operating voltage, 17,000~rpm rated vibration speed, 72~mA operating current, and 0.93~G typical normalised amplitude.  The motor was suspended vertically above 
room 4 of the setup (Fig.~\ref{setup}a)  so that  only 2~mm of its body is immersed into the water (Fig.~\ref{vibrationcorridor}a). We conducted eight experiments, in each experiment a leech was placed in the standard position indicated by a star in Fig.~\ref{setup}a. A representative time-snapshots of a typical experiment are shown in 
Fig.~\ref{vibrationcorridor}b. A leech explores terminal rooms 1 and 8 with higher frequency, yet sometimes visit room 4 and adjacent rooms. 'Sometimes' is quantified in the graph Fig.~\ref{vibrationcorridor}c. Position of a generator of water waves in room 4 causes flattening of the rooms explorations frequencies: compare Fig.~\ref{statistics}a and Fig.~\ref{vibrationcorridor}c. The room 4, where the vibrator was position, did not get any preferential visits from leech, possibly because actual configuration of waves strongly affected  by walls of the template and thus it was difficult for a leech to pinpoint exact location of the vibration source.  Experiments with small (7$\times$7~cm) maze with  a vibrating motor placed in the central chamber were inconclusive. Leeches exploring the maze were approaching the source of vibration in all six experiments, see two exemplary time-snapshots in Fig.~\ref{mazevibration}, however, after almost closed contact with the vibrating motor they were moving away. 

\begin{figure}[!tbp]
\centering
\subfigure{}{\includegraphics[width=0.49\textwidth]{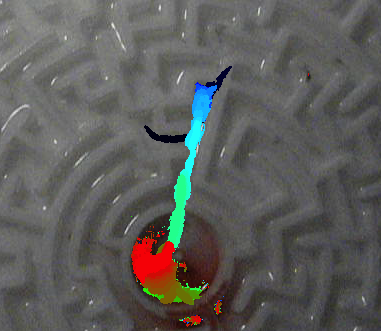}}
\subfigure{}{\includegraphics[width=0.49\textwidth]{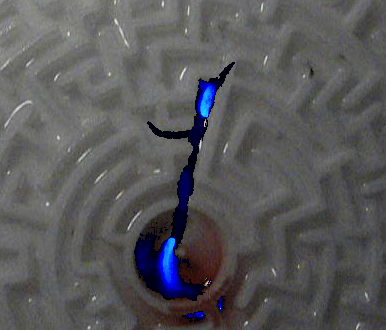}}
\caption{Exemplary images of a leech moving to a central chamber of the maze, where 0.25~ml of fresh blood were placed. (a)~Time-snapshots. (b)~Activity-snapshots: the longer leech stays in a site the brighter/lighter is a colour of pixels at this site.  }
\label{bloodmaze}
\end{figure}

There are no definite proofs that a blood \emph{per se} attracts leeches; leeches can be fed with with glucose, galactose and some other sugars if NaCl is present in the feeding solution~\cite{galun1966chemical}. It is reported, that exposure to an artificial blood actually reduces chances of swimming activity in leeches~\cite{brodfuehrer2006modification}. Nevertheless, we have done ten scoping experiments. In each experiment c. 0.25~ml of human blood (drawn from the ring finger of a healthy male individual using Unistick 3 Comfort safety lancets) was placed in the central chamber of a larger (15~cm in diameter) maze, printed from  nylon. Immediately after this a leech was introduce to the maze. Leeches initially positioned further than 5~cm away from the central chamber explored the maze disorderly without showing any preferential movement towards the source of blood. 

In three experiments leeches were placed at 5~cm distance from the source of blood, after 2-3~min being in a still mode they moved towards the central chamber. One such experiment is illustrated in Fig.~\ref{bloodmaze}.  Temporal movement is straightforward (Fig.~\ref{bloodmaze}a): at the beginning of the experiment the leech crawled towards the central chamber.  When presented with a choice of turning right and heading straight, the leach explored the space (you see the leech turning right in (Fig.~\ref{bloodmaze}a) but then continues propagation towards the central chamber. Activity-snapshots give us more insight onto modes of the leech's behaviour (Fig.~\ref{bloodmaze}b): the leech is in still till mode at the beginning of the experiment (bright, almost white blob in  Fig.~\ref{bloodmaze}a), then  hesitates a bit before entering the main chamber (small bright blob), and then switches into the still inside the central chamber, surrounded by fresh human blood slightly diluted by water in the chamber. 

Returning back to our original experiments on exploration  of the row of rooms we must highlight that our interpretation put a key role on tactile sensors and mechanical stimulation. There is indeed a chance a leech deposits some chemical substances while propagating along the corridor. More substances could be deposited at the end of the corridor due to leech spending more time their while turning her body to move in the opposite direct. It might be that excessive concentration of deposited substances attract the leech. Said we did not find any evidence in the published literature that 
leeches behave as e.g. ants do, i.e. by following trails of their own pheromones. Therefore we left verification of the 'chemical deposits' hypothesis to further studies.

\section{Conclusion}
\label{discussion}

When placed in geometrically constrained spaces  leeches  \emph{Hirudo verbana} exhibits three key behavioural modes of behaviour: still, crawling and exploration (swimming mode was rarely observed in water layers  5~mm or less). Switching between still and active (crawling and exploration) modes happens when time limit in current mode elapses. The leech goes into exploration mode when it encounters a mechanical obstacle. A probability of returning from exploratory mode to crawling mode decreases proportionally to a distance from last mechanical obstacle. When placed in a corridor with a raw of rooms on one side the leech explores rooms near end of the corridor with higher probability and rooms near the centre of the corridor with lower probability. Thus, answer to our toy question `Where is the best place to hide from a leech?' would `In the room in the middle of the corridor'. Based on the results of our laboratory experiments we formalised behaviour of a leech in terms of probabilistic finite state machines with binary inputs. The structure of the machines could be used as a blueprint in designing of future leech-robots. 

In our scoping experiments with leeches in mazes we found that leeches display slight preferential directions of movements towards sources of blood and sources of vibration. However, experiments were inconclusive and more work is necessary towards establishing exact role of tactical behaviour of leeches during navigation in complex environments. Chances are that vibration and temperature gradients do not develop well in complex mazes due to low sound and thermal insulation properties of the maze walls. Without well developed gradients it might be difficult for a leech to find exact path towards a source of vibration or high temperature. Said that we believe even the debatable results obtained in the scoping experiments indicate that leeches are very promising prototypes of future flexible amphibian rescue robots. Such robots will adopt all four important traits of leeches locomotion~\cite{friesen2007leech}: hormonal control of movement initiation expressed in probabilities of spontaneous transitions between still and active states, effective control movement achieved via integration of optical, mechanical and chemical sensory feedback with central control circuits,  
autonomous inter-segmental coordination and adaptive decision-making.

\end{document}